\documentclass[manuscript,12pt]{elsarticle}
\usepackage{graphicx}
\usepackage{colordvi}
\journal{J. Phys. B}
\begin{document}

\begin{frontmatter}
\title{Oscillations of a quasi-one-dimensional dipolar supersolid}
\author[a,b]{B. Kh. Turmanov}
\author[a]{B. B. Baizakov\corref{*}}
\ead{baizakov@uzsci.net} \cortext[*]{Corresponding author.}
\author[a,b]{F. Kh. Abdullaev}
\author[c]{M. Salerno}

\address[a]{Physical-Technical Institute, Uzbek Academy of Sciences, 100084, Tashkent, Uzbekistan}
\address[b]{Department of Theoretical Physics, National University of Uzbekistan, 100174, Tashkent, Uzbekistan}
\address[c]{Dipartimento di Fisica E.R. Caianiello, and INFN Gruppo
Collegato di Salerno, \\ Universita di Salerno, Via Giovanni Paolo
II, 84084 Fisciano, Salerno, Italy}

\begin{abstract}
The properties of a supersolid state (SS) in quasi-one-dimensional
dipolar Bose-Einstein condensate is studied, considering two
possible mechanisms of realization - due to repulsive three-body
atomic interactions and quantum fluctuations in the framework of the
Lee-Huang-Yang (LHY) theory. The role of both mechanisms in the
formation of SS properties has been emphasized. The proposed
theoretical model, based on minimization of the energy functional,
allows evaluating the amplitude of the SS for an arbitrary set of
parameters in the extended Gross-Pitaevskii equation (eGPE). To
explore the dynamics of the SS first we numerically construct its
ground state in different settings, including periodic boundary
conditions, box-like trap and parabolic potential, then impose a
perturbation. In oscillations of the perturbed supersolid we observe
the key manifestation of SS, namely the free flow of the superfluid
fraction through the crystalline component of the system. Two
distinct oscillation frequencies of the supersolid associated with
the superfluid fraction and crystalline components of the wave
function are identified from numerical simulations of the eGPE.
\end{abstract}


\end{frontmatter}

\section{Introduction}

The existence of an unusual state of matter, called {\it
supersolid}, was theoretically predicted more than half a century
ago (see review articles \cite{boninsegni2012,balibar2010}). In this
state, matter exhibits both the crystalline order and superfluidity
at the same time. More definitely, some fraction of the total mass
is rigid as a crystal, while the rest flows through the system
freely, without dissipation. The last property is considered to be
the key signature of the supersolid state (SS) and its experimental
observation has been the main objective pursued over the years of
research. To date, there are some experimental pieces of evidence
for this phenomenon in solid He, but the theoretical interpretation
of observed properties remains inconclusive, as summarized in
\cite{prokof'ev2007,yukalov2020}. Recent studies have discovered,
that quantum gases with strong dipole-dipole atomic interactions can
be another medium, where the SS clearly shows up
\cite{tanzi2019a,bottcher2019,chomaz2019,tanzi2019b}. The advantage
here is that the quantum gases appear to be a highly controllable
coherent system. The physical mechanism behind the emerging SS in
this system is the delicate balance between the long-range dipolar
and short-range contact interactions among the atoms of the gas. The
stabilizing effect of quantum fluctuations is essential for the
existence of quantum droplets and dipolar supersolids (see review
articles \cite{luo2021,guo2021,bottcher2021}). The short-range
interactions originate from two-body and three-body atomic
collisions, characterized by the $s$-wave scattering length $a_s$,
while the quantum fluctuation effects are described in the local
density approximation by the Lee-Huang-Yang (LHY) theory
\cite{lee1957,petrov2015}. The quantum fluctuations and elastic
three-body atomic interactions were shown to play similar roles in
the process of stabilization of self-bound quantum droplets
\cite{gautam2019}. It is natural to suggest, that they play similar
roles in stabilizing the supersolids as well. Formation of an array
of quantum droplets in dipolar Bose-Einstein condensates (BEC),
stabilized solely by the conservative three-body atomic
interactions, was theoretically demonstrated
\cite{bisset2015,xi2016}. The model was able to reproduce
experimental observations, confirming the insignificance of the
three-body losses for the relevant time scales. A new kind of
dipolar quantum droplets, which can be made of strongly interacting
bosons, was reported in \cite{oldziejewski2020}. In the strongly
interacting Tonks-Girardeau regime, the governing equation of Ref.
\cite{oldziejewski2020} reduces to the nonlocal Gross-Pitaevskii
equation (GPE) with quintic nonlinearity, previously studied in the
context of flat-top localized states in quasi-1D dipolar condensates
\cite{kolomeiski2000,baizakov2009}, while in the weakly interacting
limit it reduces to usual nonlocal GPE with cubic nonlinearity
\cite{cuevas2009,baizakov2019}. In a recent study \cite{ota2020},
the authors have derived the modified GPE, taking into account the
next order (beyond LHY) quantum fluctuation effects. The additional
term in the energy functional has a cubic dependence on the gas
density, i.e. formally similar to the effect of attractive
three-body atomic interactions. A conservative quintic nonlinear
term whose contribution may be of the same order as of the LHY
effect, may arise also from the expansion of the well known quasi-1D
nonpolynomial Schr\"odinger equation (NPSE)
\cite{salasnich2002,gligoric2009}. It follows from the above
arguments that the extended Gross-Pitaevskii equation (eGPE) with
cubic-quintic nonlinearity and usual LHY term can be a more adequate
framework for describing the SS in dipolar quantum gases, therefore
it is adopted in the present study.

The basic properties of condensed matter systems can be revealed
from the analysis of their excitation spectra. The spectrum of
elementary excitations in dipolar quantum gases of $^{162}$Dy and
$^{166}$Er, confined to a harmonic trap, was studied in
\cite{tanzi2019b,natale2019}. It was experimentally demonstrated,
that in the regime of BEC the system exhibits ordinary quadrupole
oscillations with a single frequency, while in the supersolid regime
two distinct frequencies arise, corresponding to oscillations of the
superfluid and crystalline phases. An important result of these
works is that the two-frequency response of a supersolid, originally
predicted for infinite-size systems, was confirmed for finite-size
trapped quantum gases. The possibility of dynamical and simultaneous
excitation of the roton and maxon modes in Rydberg-dressed
condensates through interaction quenches was reported
\cite{mccormack2020}. Supersolid behavior of a dipolar BEC was
addressed through numerical simulations of the Bogolyubov - de
Gennes equations in a tubular periodic confinement
\cite{roccuzo2019}. In this work, the coherence of the system across
the phase transitions BEC - supersolid - isolated quantum droplets
was analyzed. Description of the ground state and collective
excitations of a supersolid in dipolar quantum gases, based on the
variational approach, was presented in
\cite{blakie2020a,blakie2020b}. A comprehensive introduction to the
nonlocal GPE with the contribution of quantum fluctuation effects in
a quasi-one-dimensional setting can be found in \cite{edmonds2020}.
Despite impressive experimental and theoretical progress in
exploring the SS in dipolar quantum gases, some static and dynamic
aspects of trapped supersolids remain less studied.

Our objective in this work consists in exploring the conditions,
which can support a robust SS in a quasi-one-dimensional dipolar
Bose-Einstein condensate with competing long-range dipole-dipole and
short-range contact atomic interactions, in the presence of quantum
fluctuation effects. A theoretical model will be developed to
estimate the amplitude of the emerging SS. It should be noted that
the amplitude of the SS is directly linked to the experimentally
measurable quantity, called density contrast
\cite{petter2020,chomaz2020}. Therefore the proposed model can be
useful for the analysis and interpretation of experimental results.
Besides, we numerically investigate the dynamics of a supersolid,
subject to a weak perturbation, to demonstrate its key property,
namely the free flow of the superfluid fraction of the quantum gas
through its crystalline fraction. When perturbed, the crystalline
and superfluid components of the system perform undamped
oscillations with different frequencies
\cite{tanzi2019b,natale2019,guo2019}, which is the hallmark property
of supersolids. In our numerical simulations by recording the
time-dependence of the width of the oscillating supersolid, and
Fourier-analyzing the obtained data, we identify these two distinct
frequencies.

The paper is structured as follows. In the next Sec. II we present
the main equations, the spectrum of elementary excitations, and the
variational approach for the amplitude of the supersolid. Sec. III
is devoted to SS dynamics subject to perturbation. In Sec. IV we
summarize our findings.

\section{The model and main equations}

The model is based on a normalized quasi-1D eGPE, derived from its
full three-dimensional version (for details of 3D to 1D reduction
see \cite{edmonds2020,turmanov2020})
\begin{eqnarray}\label{gpe}
i\psi_t + \frac{1}{2}\psi_{xx} - V(x)\psi + g |\psi|^2 \psi + \gamma
|\psi|^3 \psi + p|\psi|^4\psi + & & \nonumber \\
  g_{dd}\; \psi(x,t)\int_{-\infty }^{+\infty} R(|x-x'|)\ |\psi
  (x',t)|^2 \,dx' =0,
\end{eqnarray}
where $\psi(x,t)$ is the mean-field wave function of the condensate,
$V(x)$ is the trap potential, $g, p, g_{dd}$ are the coefficients of
two-body, three-body and dipole-dipole atomic interactions,
respectively. The parameter $\gamma$ quantifies the contribution of
quantum fluctuations,\\ $R(x) = \sqrt{\pi} \exp{\left(x^2\right)}
\left( 1+2 x^2\right) \textrm{erfc} \left(|x| \right) - 2|x| $ is
the response function which characterizes the degree of nonlocality
of the medium \cite{sinha2007}. The Eq. (\ref{gpe}) is obtained for
the condensate, which is tightly confined in the radial direction,
with the transverse and axial trap frequencies satisfying the
condition $\omega_{\bot} \gg \omega_x$. Under such confinement, the
condensate acquires a highly elongated form in the axial direction
commonly termed as cigar-shaped condensate. Below we consider
different types of confinement, namely a tubular geometry with
periodic boundary conditions, mimicking the infinite system,
box-like potential with soft walls, and parabolic trap potential.
The dimensionless time, space, and wave function in Eq. (\ref{gpe})
are linked to corresponding physical quantities as follows $t
\rightarrow \omega_{\bot} t$, $x \rightarrow x/a_{\bot}$, $\psi
\rightarrow \sqrt{2|a_{bg}|}\psi$, with $a_{bg}$ being the
background value of the $s$-wave scattering length, $a_{\bot} =
\sqrt{\hbar/m \omega_{\bot}}$ is the radial harmonic oscillator
length, $m$ is the atomic mass. The dimensionless coefficients of
Eq.~(\ref{gpe}) are expressed through the actual parameters of the
system as follows $g = - a_s/|a_{bg}|$, \\ $p=-g_2
m^2\omega_{\bot}/12 \pi^2 \hbar^3 a_s^2$, $g_{dd}~=~-~\mu_0
\mu^2/(4\pi\hbar^2 |a_{bg}|/m)$, where $g_2$ characterizes the
strength of three-body atomic interactions, $\mu_0, \mu$ are the
permeability of vacuum and atomic magnetic moment, respectively. In
these notations $g>0$ and $g_{dd}<0$, $p<0$ corresponds to
attractive contact interactions competing with the repulsive dipolar
and three-body terms. The contribution of quantum fluctuations is
characterized by $\gamma = 128\,a_s (1+1.5
\,\varepsilon_{dd}^2)/15\sqrt{2} \pi a_{\bot}$, where
$\varepsilon_{dd} = m \mu_0 \mu^2/12 \pi \hbar^2 |a_s|$ is the ratio
between the dipolar and $s$-wave scattering lengths. The LHY energy
correction in dipolar condensates leads to a repulsive term $\sim
\gamma\,n_{1D}^{3/2}$ in the eGPE (\ref{gpe}) provided that the
one-dimensional density of the gas $n_{1D}$ is sufficiently large
($n_{1D} a_s > 0.6$) \cite{edler2017}. It is relevant to mention
that the coefficient $p$ in Eq. (\ref{gpe}) may also be associated
with the term of quintic nonlinearity arising from expansion of the
NPSE \cite{salasnich2002,gligoric2009}, or the beyond LHY
description of quantum fluctuation effects \cite{ota2020}.

\subsection{Dispersion relation for elementary excitations}

The spatial period of emerging density modulations in a dipolar
condensate, which can transform in suitable conditions into SS, can
be defined from the dispersion curve of elementary excitations. To
derive the dispersion relation we consider the ground state wave
function of the form
\begin{equation}
\psi_0= \sqrt{n_0} e^{i \theta t},
\end{equation}
where $n_0$ is the constant background density, $\theta$ is the
uniform phase
\begin{equation}\label{phase}
\theta=n_0\left(g + p n_0 + \gamma {n_0}^{1/2} + g_{dd}
\int_{-\infty}^{\infty}R(|x|) dx\right).
\end{equation}
Next, we introduce weak perturbation to the ground state $\psi(x,t)=
\psi_0 + \delta\psi(x,t)$, assuming $\delta\psi(x,t) = \eta(x,t)
e^{i \theta t}$ and $|\eta(x,t)|^2 \ll n_0$. Substituting
$\psi(x,t)$ into Eq. (\ref{gpe}) and linearizing around
$\psi_0(x,t)$ one obtains
\begin{eqnarray}
i \eta_t + \frac{1}{2} \eta_{xx} + n_0\left(g+2 p n_0+\frac{3}{2}
\gamma n_0^{1/2}\right)(\eta(x,t)+\eta^{\ast}(x,t)) + & &  \nonumber \\
g_{dd} n_0\int_{-\infty}^{\infty}
R(x-x')(\eta(x',t)+\eta^{\ast}(x',t))dx' &=&0.
\end{eqnarray}
Splitting the perturbation field into real and imaginary parts
$\eta(x,t)=u(x,t)+ i v(x,t)$, and taking the Fourier transform of
the resulting equation, yields the following coupled system for the
transformed components
\begin{eqnarray}
&& \dot{\tilde u}(k,t)= \frac 12 k^2 \tilde v(k,t), \\
&& \dot{\tilde v}(k,t)=-\left[\frac{1}{2} k^2 - 2 n_0 \left(g + 2 p
n_0 + \frac{3}{2} \gamma n_0^{1/2} + g_{dd} \tilde
R(k)\right)\right] \tilde u(k,t), \nonumber
\end{eqnarray}
where the overdot denotes the derivative with respect to time, while
the tilde stands for the Fourier transform $\tilde{f}(k)=
\int_{-\infty}^{\infty} f(x) e^{i k x} dx$, and $\tilde{R}(k)$ is
given~by
\begin{equation}\label{rf}
\tilde R (k)= 2\left[1+ \frac{k^2}{4} e^{k^2/4} {\rm
Ei}\left(-\frac{k^2}{4}\right)\right],
\end{equation}
with ${\rm Ei}(z)$ being the exponential integral function
\cite{abramowitz}. Notice that the above  equations are equivalent
to a single harmonic oscillator equation
\begin{equation}\label{ho}
\ddot{\tilde u} + \Omega^2(k) \tilde u = 0,
\end{equation}
with the frequency
\begin{equation}\label{Omega}
\Omega(k) =  k \left(\frac{k^2}{4} - n_0 (g + 2 p n_0 +
\frac{3}{2}\gamma n_0^{1/2} + g_{dd} \tilde{R}(k) ) \right)^{1/2}.
\end{equation}
For small values of the wave vector $k$ the above frequency defines
the energy spectrum $E(k)=v_s k$ of the elementary excitations
obtained from the Bogoliubov-de Gennes analysis with a sound
velocity
\begin{equation}\label{vs}
v_s=\sqrt{-n_0 \left(g + 2 p n_0 + \frac{3}{2}\gamma n_0^{1/2} +
g_{dd} \tilde{R}(k)\right)}.
\end{equation}
For larger values of $k$ and suitable parameters the dispersion
relation (\ref{Omega}) features a local minimum at $k=k_r$,
corresponding to quasi-particles known as {\it rotons}, whose
existence in dipolar condensates was theoretically
predicted~\cite{santos2003} and experimentally confirmed
\cite{chomaz2018,petter2019}. Although initially the roton was
linked to local vorticity in superfluid $^4$He~\cite{landau1941},
nowadays it is viewed as the precursor of a crystallization
instability \cite{nozieres2004}.

\begin{figure}[htb]
\centerline{\includegraphics[width=6cm,height=5cm,clip]{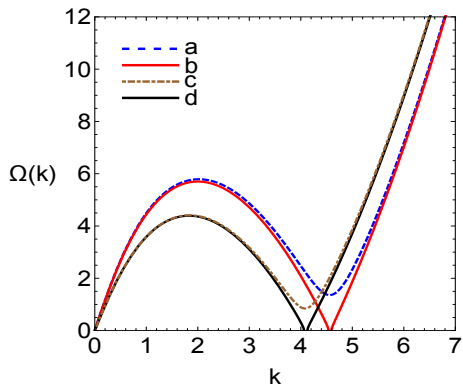}}
\caption{The excitation spectrum according to Eq. (\ref{Omega}) for
$n_0=12$, $g=1$ and different values of the dipolar ($g_{dd}$),
quintic ($p$) and LHY ($\gamma$) coefficients: (a) $g_{dd}=-2$,
$p=0$, $\gamma=0$; (b) $g_{dd}=-1.974$, $p=0$, $\gamma=0$; (c)
$g_{dd}=-1.41$, $p=-0.007$, $\gamma=-0.002$; (d) $g_{dd}=-1.41$,
$p=-0.007$, $\gamma=-0.0013$. The black and red curves touch the
zero axis at some value of $k_r=4.1$ and $k_r=4.57$ respectively,
thus $\Omega(k)$ becomes imaginary. } \label{fig1}
\end{figure}

Figure \ref{fig1} illustrates the dispersion relation (\ref{Omega})
for different sets of parameter values. The real and positive values
of the frequency $\Omega(k)>0$ (blue dashed and brown dot-dashed
lines in Fig. \ref{fig1}) signify that the density modulations in
the condensate are of standing wave type. A small energy gap near $k
\sim k_r$ implies that the roton quasi-particles are more
effectively generated in the condensate compared to other types of
excitations \cite{turmanov2020}. The supersolid emerges as the
frequency (\ref{Omega}) becomes imaginary (red and black solid lines
in Fig. \ref{fig1}) and roton instability sets in
\cite{giovanazzi2004}. The steady growth of the amplitude of density
waves, caused by the roton instability, is counterbalanced by the
repulsive LHY term and/or three-body atomic scattering effects. It
should be stressed that the balance solely between the attractive
two-body ($g>0$) and repulsive dipolar ($g_{dd}<0$) interactions is
unstable (corresponding to the curve (a) in Fig. \ref{fig1}).
Therefore, the presence of repulsive LHY and/or three-body
interaction terms is essential for the existence of stable SS in
dipolar condensates. Either of these two higher-order nonlinear
terms or a combination of them can stabilize the supersolid,
producing very similar stationary density profiles. The spatial
period of the supersolid represents its most important
characteristic which is defined by the wave vector of rotons
$\lambda = 2\pi/k_r$. Another important parameter is the amplitude
of the SS, expressed through its field contrast as $a_f={\rm
max}(|\psi(x)|)-{\rm min}(|\psi(x)|)$, where ${\rm max}$ (${\rm
min}$) denotes the maximum (minimum) over the space.
Below we study the existence of SS in dipolar quantum gases and its
dynamics employing the variational approach and numerical
integrations of the eGPE~(\ref{gpe}).

\subsection{Variational approach for supersolids in dipolar quantum gases}

It was pointed out earlier that supersolids in dipolar condensates
can exist due to the stabilizing effect of repulsive quantum
fluctuations and/or three-body atomic interactions. The existence
conditions and some characteristics of the supersolid can be
evaluated using the variational approach. A variational theory based
on minimization of the GPE energy functional was developed for the
former \cite{blakie2020a,blakie2020b} and latter \cite{lu2015}
cases, using the cosine modulated trial function $\psi(x) =
\sqrt{n}\left(\cos \theta +\sqrt{2}\sin \theta \cos(k x)\right)$,
with $n$ being the average linear density of the condensate. The
variational parameters $\theta$ and $k$ characterize the amplitude
and period of the density modulations, respectively. This ansatz is
valid for weakly modulated density waves in the absence of trapping
potential.

Below we develop a variational approach for supersolids in dipolar
quantum gases using an alternative trial function
\begin{equation}\label{ansatz}
\psi= A + a \cos(k x),
\end{equation}
where the variational parameters ($A,a,k$) are introduced, meaning
the amplitude of the background, amplitude and wave vector of
density modulations, respectively. The normalization to the average
linear density of the condensate
\begin{equation}\label{norm}
n=\lim_{L\to\infty}\left(L^{-1}\int_{-L/2}^{L/2} {|\psi(x,t)|^2}
dx\right) = \frac{a^2}{2} +A^2
\end{equation}
is adopted. Similar (sine-modulated) trial function was previously
employed in Ref. \cite{chomaz2020}. These trial functions are
suitable when the axial confinement is absent ($\alpha=0$). By
substitution of the ansatz (\ref{ansatz}) into the energy density,
corresponding to Eq. (\ref{gpe}) without the harmonic trap
\begin{equation}
{\cal E}=\frac{1}{2}|\psi_{x}|^2-\frac{g}{2}|\psi|^4 -
\frac{2}{5}\gamma|\psi|^5-\frac{p}{3}|\psi|^6-
\frac{g_{dd}}{2}|\psi|^2\int\limits_{-\infty}^{+\infty}R(|x-x^\prime|)\
|\psi(x^\prime,t)|^2 \ dx^\prime,
\end{equation}
and integration over the space variable
$E=\lim_{L\to\infty}\left(L^{-1}\int_{-L/2}^{L/2} {\cal E}
dx\right)$, one obtains the following GP energy functional
\begin{eqnarray}\label{energy}
E=\frac{a^2 k^2}{4} - \frac{g}{16} \left(8 n^2 +16 a^2 n - 7 a^4
\right) -\frac{\gamma}{20} \sqrt{n-\frac{a^2}{2}} \left(8 n^2 +32
a^2 n -3 a^4 \right) - \qquad && \nonumber \\
\frac{p}{24} \left( 8 n^3 +48 a^2 n^2 -9 a^4 n - 6 a^6 \right)
-g_{dd}\left[ n^2 + a^2 \left(n - \frac{a^2}{2} \right) \tilde{R}(k)
+ \frac{a^4}{16} \tilde{R}(2k) \right]. \ &&
\end{eqnarray}
The response function in the Fourier space $\tilde{R}(k)$, defined
by Eq. (\ref{rf}), has the following properties
\begin{equation}\label{dr}
\tilde{R}(0)=2, \quad \tilde{R}'(k)=\frac{k^2 + 4}{2
k}\tilde{R}(k)-\frac{4}{k},
\end{equation}
where the prime stands for the derivative with respect to $k$. Using
these relations the expression for the amplitude of density
modulations $a$ can be derived from minimization of the energy
($dE/dk=0$)
\begin{equation}\label{a}
a = 2\left(\frac{k^2 +8 g_{dd} n - g_{dd} n (k^2
+4)\tilde{R}(k)}{g_{dd} ((k^2 + 1)\tilde{R}(2 k)- 2 (k^2 +4
)\tilde{R}(k) + 14)} \right)^{1/2},
\end{equation}
and the vanishing of the roton gap ($\Omega(k)=0$) at some $k>0$.
The supersolid amplitude Eq. (\ref{a}) implicitly depends on the
system parameters through $k = k_r = f(g,p,\gamma,g_{dd},n)$.
Calculations according to Eq. (\ref{Omega}) and Eq. (\ref{a}) for
parameter values specified for the black line (d) in Fig. \ref{fig1}
gives for the roton minimum $k_r = 4.1$, and SS amplitude $a=0.4$ at
average linear density $n=12.12$, which is in good agreement with
the results of eGPE ($a=0.46$).

The validity of the variational Eq. (\ref{a}) is restricted to weak
(sinusoidally modulated) supersolids. When it is composed of
strongly localized spikes of density, the trial function
(\ref{ansatz}) is no longer appropriate.

\section{Dynamics of the perturbed supersolid}

The basic properties of a supersolid can be studied by observing its
collective dynamics. Oscillations of the supersolid can be induced
in different ways, such as changing the atomic interactions or
variations of the trapping potential. Below we present the results
of numerical simulations where the oscillations of the supersolid in
three different configurations have been recorded and Fourier
analyzed. The first case belongs to periodic boundary conditions,
emulating the infinite system. The second and third cases correspond
to box-like and parabolic traps respectively.

The starting point is the ground state of a SS, which is numerically
obtained by the method of Pitaevskii phenomenological damping
\cite{pitaevskii1958,choi1998}, applied to eGPE(\ref{gpe}). The
governing Eq. (\ref{gpe}) has been solved by combination of the fast
Fourier transform and fourth order Runge-Kutta procedures, using the
phenomenological damping parameter $\Lambda=-0.03$ \cite{choi1998}.
The number of Fourier modes and time step were ${\cal N}_F = 2048$
and $\Delta t = 0.0001$, respectively. The length of the integration
domain is selected to accommodate nine periods of the supersolid
$L=9\, \lambda$, with $\lambda=2\pi/k_r$, $k_r$ being the roton wave
number. The parameter values for $g, \gamma, p, g_{dd}$ in Eq.
(\ref{gpe}) correspond to situation, when the roton minimum of the
dispersion curve (\ref{Omega}) touches the zero axis (see Fig.
\ref{fig1}). The oscillations of the supersolid can be induced
either by a slight variation of the coefficient of atomic
interactions, or modification of the trap potential. The main
parameter of interest for trapped supersolid will be its time
dependent average width
\begin{equation}\label{W}
W(t)=\left(\frac{1}{{\cal N}}\int_{-L/2}^{L/2} x^2 |\psi(x,t)|^2
dx\right)^{1/2},
\end{equation}
where ${\cal N}=\int \limits^{L/2}_{-L/2} |\psi(x,t)|^2 dx$ is the
norm of the wave function, which is linked to the number of atoms in
the condensate through $N=(a_{\bot}/2 a_s){\cal N}$. For a uniform
condensate with amplitude $A$, confined to a ring trap ${\cal N} =
A^2 L$.

\subsection{Supersolid with periodic boundary conditions}

\begin{figure}[htb]
\centerline{{\large a)} \hspace{6cm} {\large b)} } \centerline{
\includegraphics[width=6cm,height=4cm,clip]{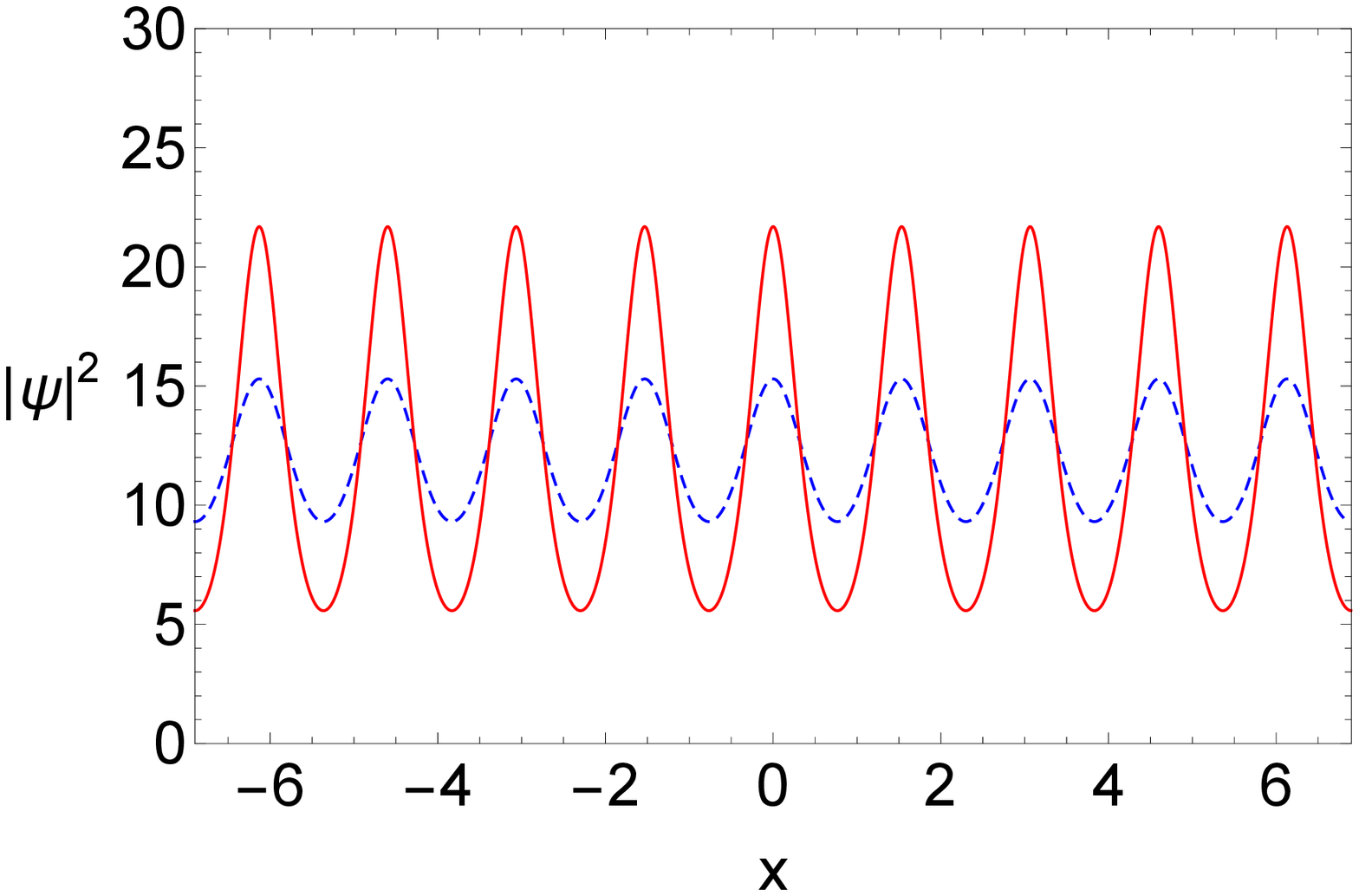}\quad
\includegraphics[width=6cm,height=4cm,clip]{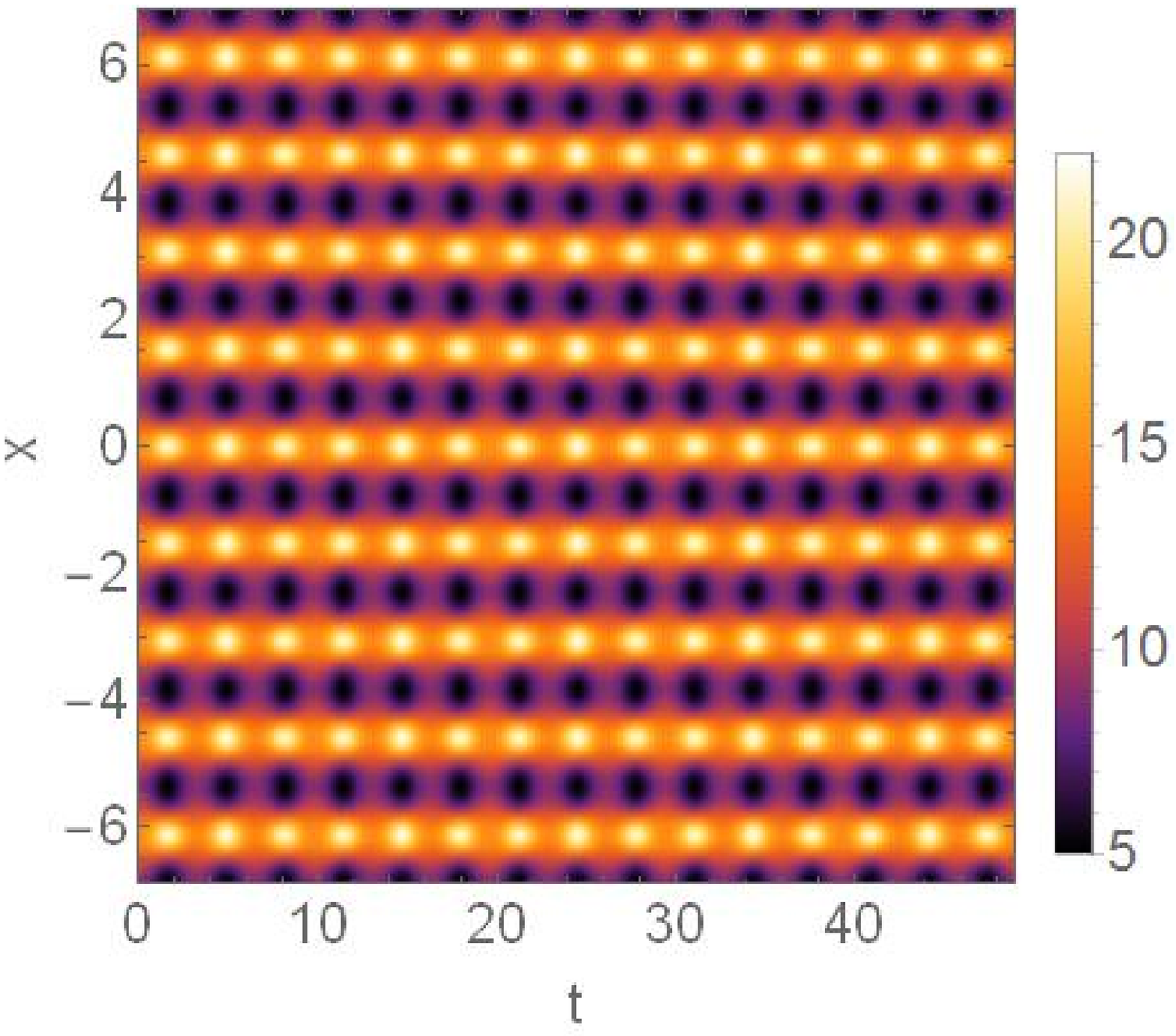}}
\caption{a) Stable SS (blue dashed line) produced by
phenomenological damping procedure, applied to eGPE (\ref{gpe}),
starting from initial condition $\psi(x)=A+a \cos(k x)$ with
$A=2\sqrt{3}$, $a=0.4$, $k=4.1$. Changing the strength of dipolar
interaction from $g_{dd}=-1.41$ to $g_{dd}=-1.38$ induces the SS
oscillations, whose wave profile is shown for $t=0$ (blue dashed
line) and $t=1.8$ (red solid line). b) The density plot
$|\psi(x,t)|^2$ showing the excitation of the SS amplitude mode
caused by variation of the dipolar coefficient $g_{dd}$ as above,
according to eGPE (\ref{gpe}). Parameter values: $g=1$, $p=-0.007$,
$\gamma=-0.001$.} \label{fig2}
\end{figure}

The majority of theoretical research concerns the infinite
supersolid, which can be emulated in numerical simulations using
periodic boundary conditions. Figure \ref{fig2} illustrates the
ground state of the supersolid and its oscillations induced by
slight variation of the strength of dipolar interactions.
It should be pointed out, that in case of periodic boundary
conditions, variation of the strength of atomic interactions leads
to excitation of the amplitude mode only, leaving the width
Eq.(\ref{W}) unaffected. Owing to a superfluid nature of the system
and equal heights of density peaks, the SS performs undamped
oscillations of its amplitude with a single frequency $\omega \simeq
1.9$ (see Fig. \ref{fig2}b).

\begin{figure}[htb]
\centerline{{\large a)} \hspace{6cm} {\large b)} } \centerline{
\includegraphics[width=6cm,height=4cm,clip]{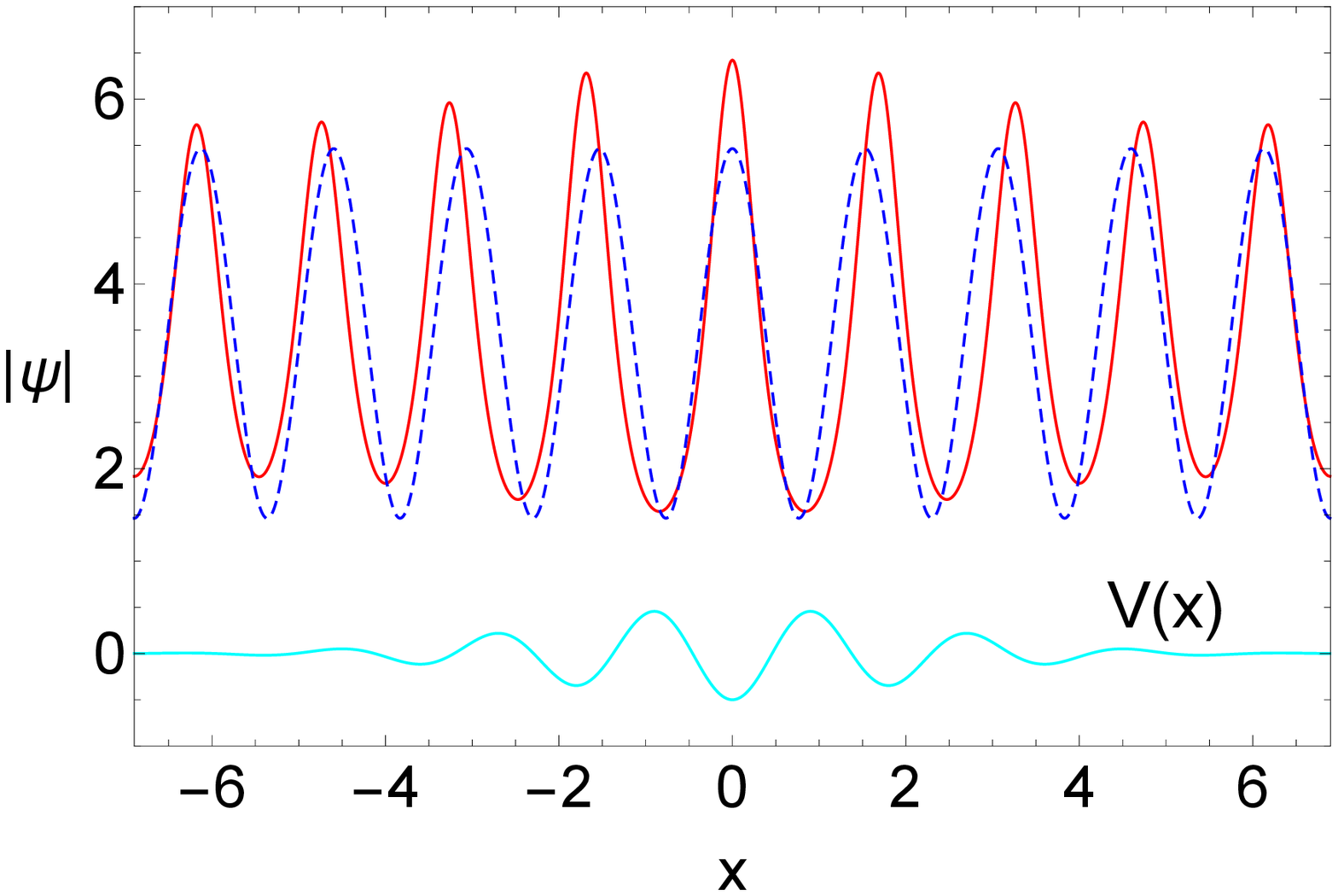}\quad
\includegraphics[width=6cm,height=4cm,clip]{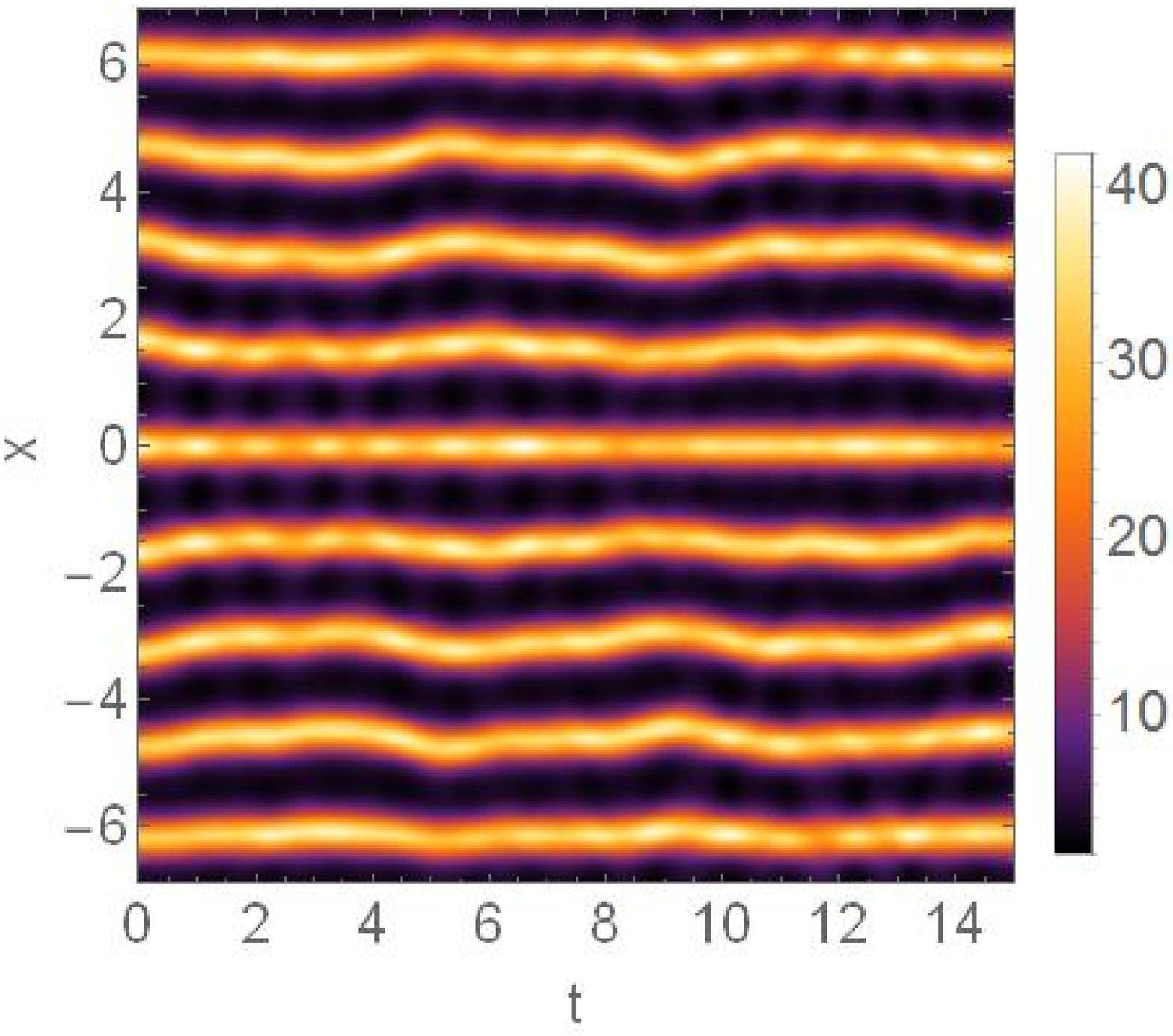}}
\centerline{{\large c)} \hspace{6cm} {\large d)}}
\centerline{\includegraphics[width=13cm,height=4cm,clip]{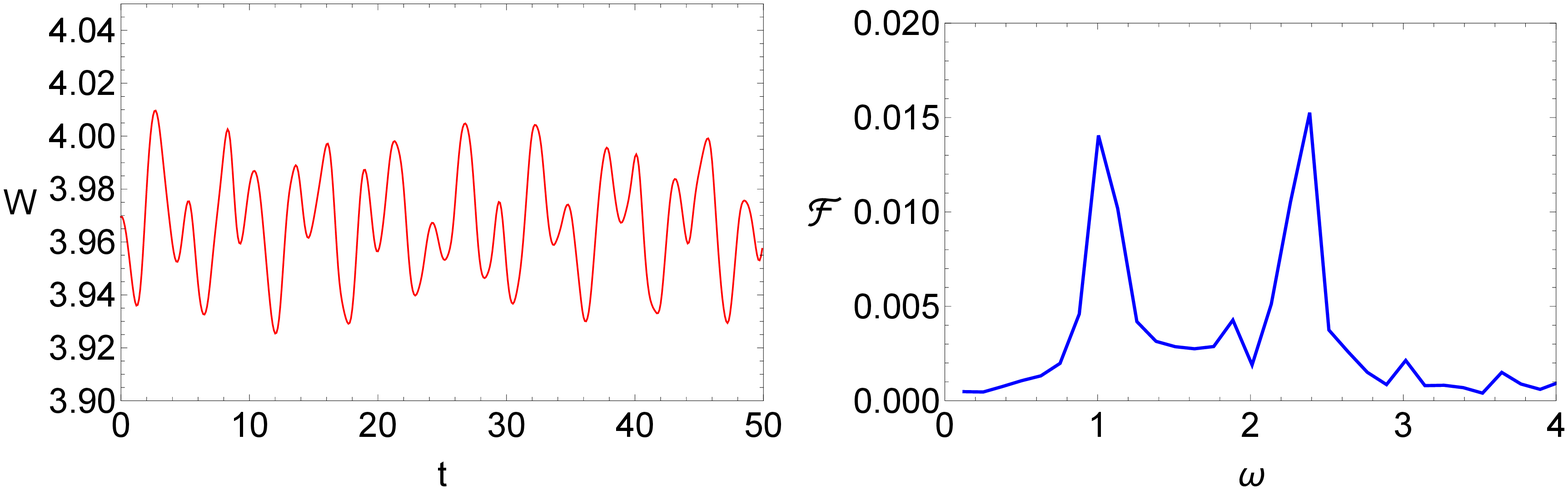}}
\caption{a) The external potential given by Eq. (\ref{pot}) (green
line) gives rise to displacement of density peaks and variation of
their heights (red solid line) as compared to the initial state
without the potential (blue dashed line). b) The density plot,
showing the oscillations of the supersolid, according to eGPE
(\ref{gpe}). c) Time variation of the effective width of the
supersolid Eq. (\ref{W}) and its Fourier transform (d). Parameter
values are the same as in Fig. \ref{fig2}, except $a=2$,
$p=-0.006$.} \label{fig3}
\end{figure}

Oscillations of the effective width can be induced using the
external potential providing a small shift of the density peaks from
their equilibrium positions, meantime not violating the periodic
boundary conditions. In numerical simulations we employ the
following potential
\begin{equation}\label{pot}
V(x)= V_0 \, {\rm cos} \left(\beta \frac{2\pi}{k_r}
x\right)\exp\left(-\frac{x^2}{w^2}\right),
\end{equation}
with parameter values $V_0=0.5$, $\beta = 2.24$, $w=3$. In addition
to the displacement of the peak positions, their heights also become
unequal. In Fig.~\ref{fig3} we show the ground state of the
supersolid in presence of the potential~(\ref{pot}), and its
dynamics caused by subsequent elimination of the potential. The
excitation of both the superfluid and crystalline modes is evident.
The Fourier analysis of the time dependent width Eq. (\ref{W})
reveals two distinct frequencies, associated with the superfluid
($\omega_s \sim 1$) and crystalline ($\omega_c \sim 2.4$) components
of the system.
If the density peaks of the supersolid become unequal due to the
external potential, a notable mixing of oscillations occurs. As a
consequence, additional small peaks arise in the Fourier spectrum
(Fig.~\ref{fig3}d).

\subsection{Supersolid in a box-like potential}

Supersolids in confined geometries exhibit more complex dynamics.
Below we consider the SS confined to a soft-wall box potential
\begin{equation}\label{box}
V(x)=\frac{V_0}{2}\left[{\rm th}\left(\frac{x-h}{w}\right) - {\rm
th}\left(\frac{x+h}{w}\right)+2\right],
\end{equation}
where the parameters $V_0, h, w$ characterize its strength, spatial
extent, and width of the transition region.

\begin{figure}[htb]
\centerline{{\large a)} \hspace{6cm} {\large b)}}
\centerline{\includegraphics[width=6cm,height=4cm,clip]{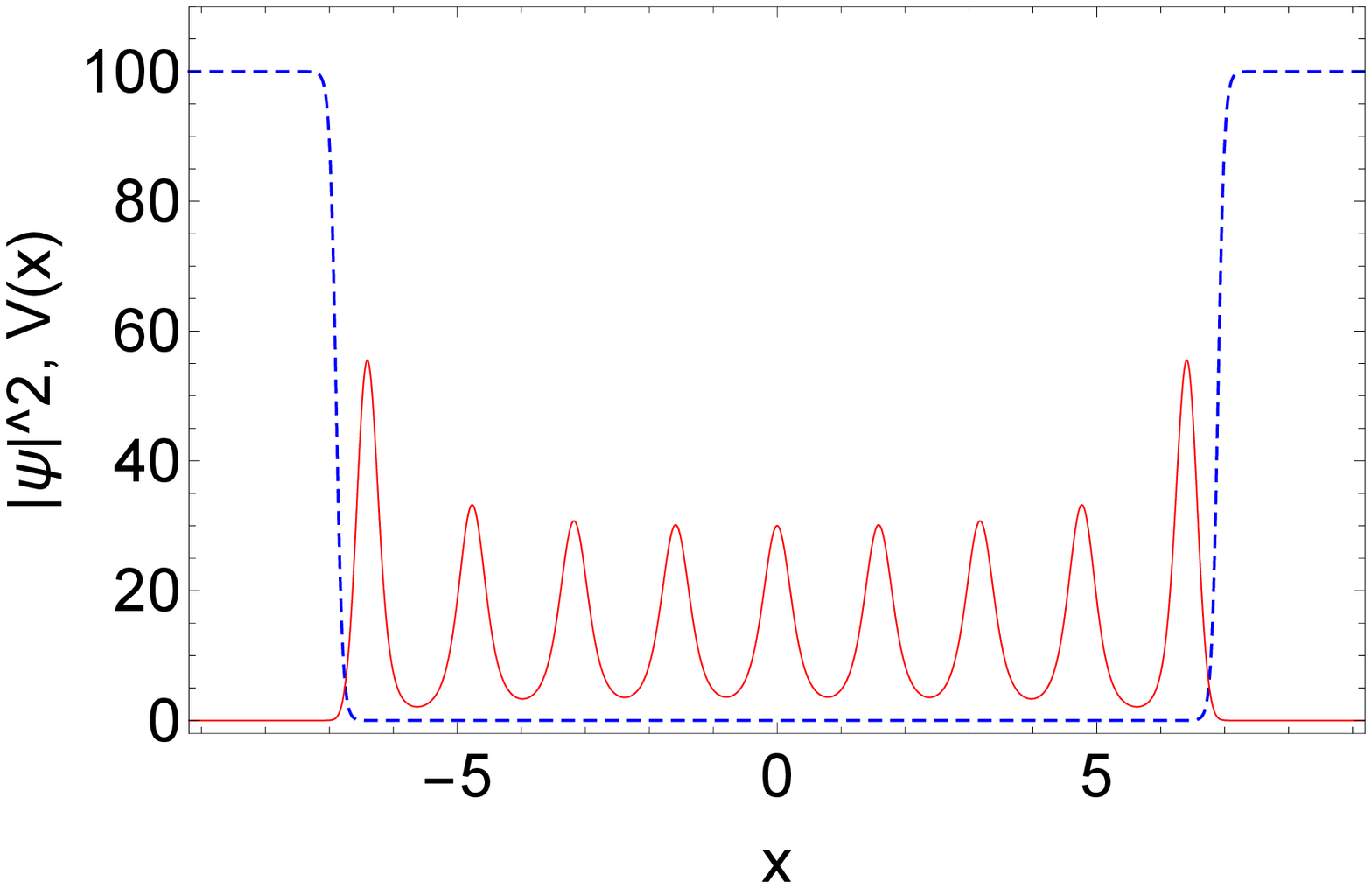}\qquad
            \includegraphics[width=6cm,height=4cm,clip]{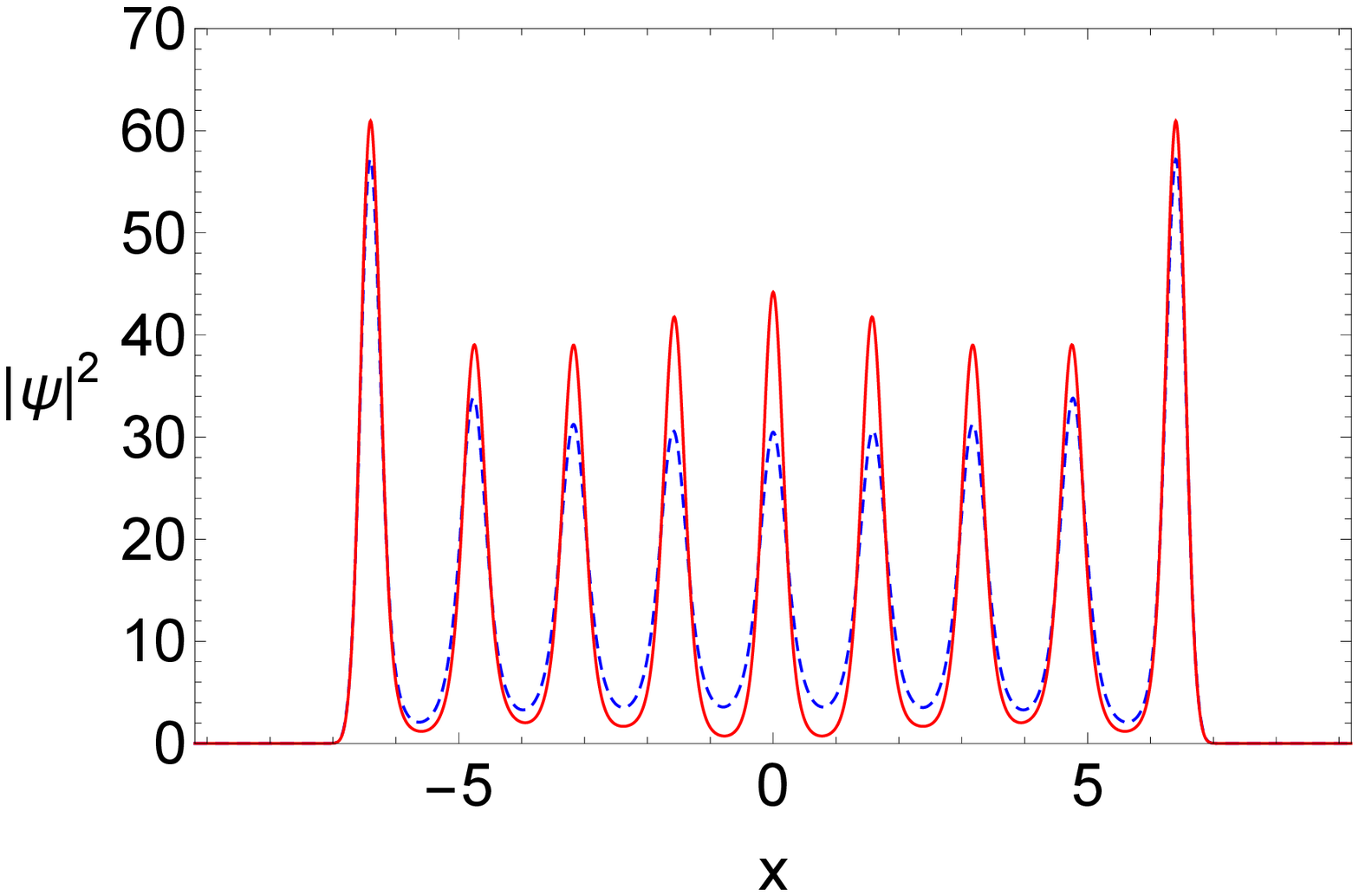}}
\centerline{{\large c)} \hspace{6cm} {\large d)}}
\centerline{\includegraphics[width=6cm,height=4cm,clip]{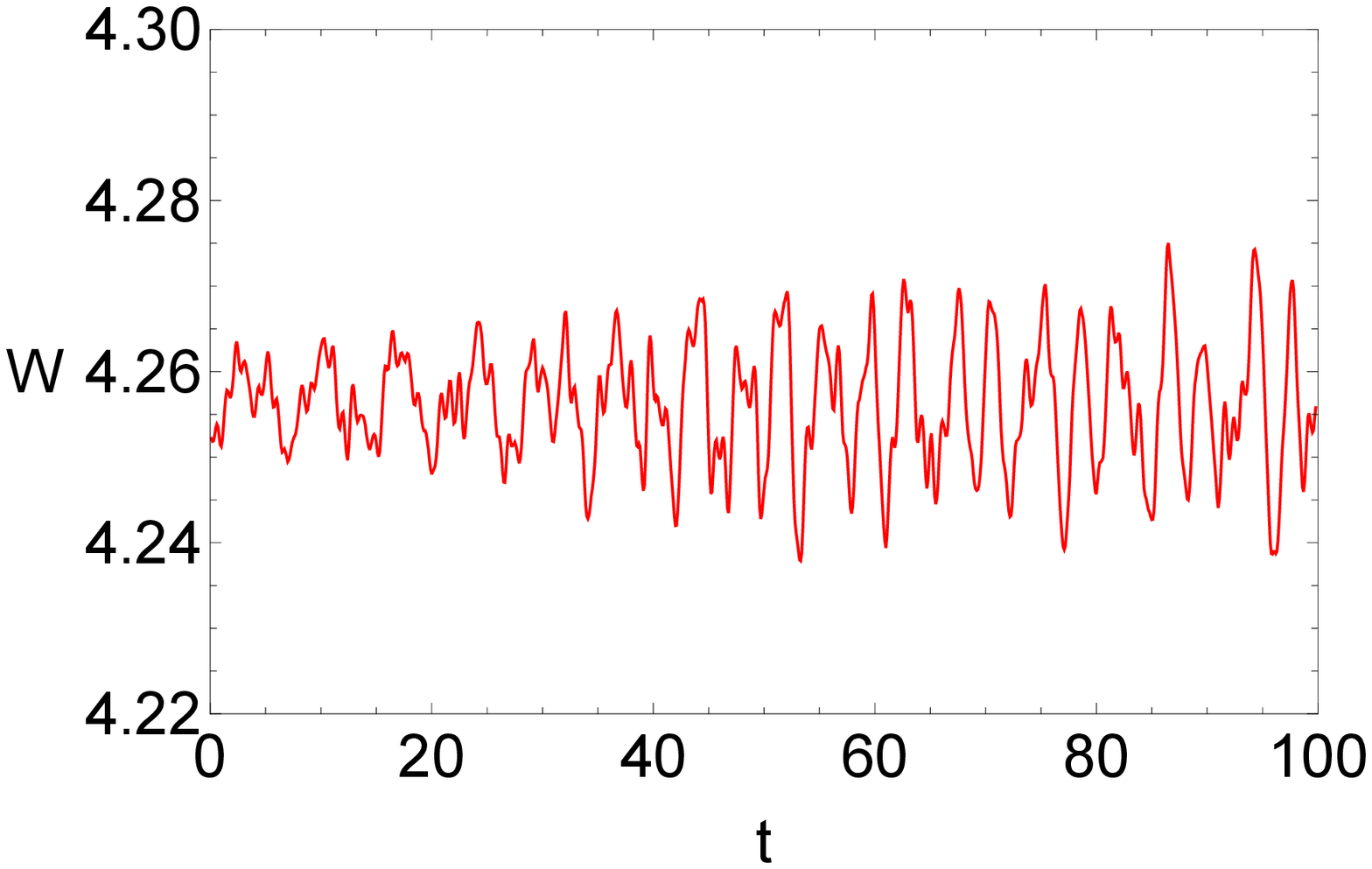}\quad
            \includegraphics[width=6cm,height=4cm,clip]{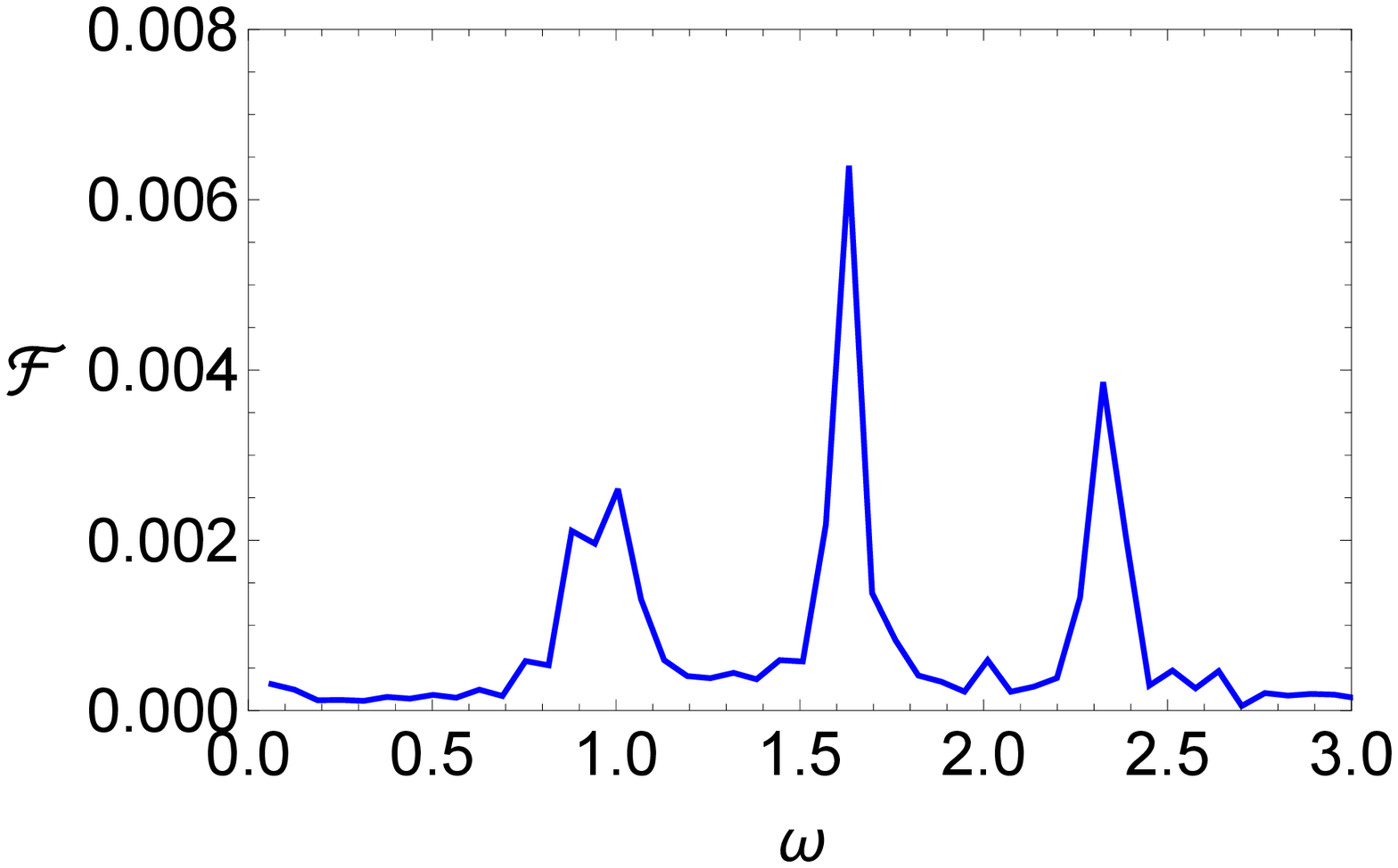}}
\caption{a) The ground state of the supersolid (red solid line) in
the box potential (\ref{box}) (blue dashed line) with $V_0 = 100$,
$w=0.1$, $h=4.5 \lambda$. b) Change of the strength of dipolar
interactions from $g_{dd}=-1.41$ to $g_{dd}=-1.27$ induces the
oscillations of the supersolid, whose profile is shown for $t=0$
(blue dashed line) and $t=1.5$ (red solid line). The central density
peaks oscillate with greater amplitude than the edge states. c) The
average width $W(t)$ performs complex dynamics. d) The Fourier
spectrum of $W(t)$ shows three main frequencies, associated with the
superfluid and crystalline components, and their superposition.
Parameter values: $g=1$, $p=-0.006$, $\gamma=-0.001$. The length of
integration domain is $L=12\lambda$.} \label{fig4}
\end{figure}

In experiments the potential (\ref{box}) can be created by laser
beams \cite{meyrath2005,navon2021}. To study oscillations of the
supersolid first we create its ground state in the box potential, as
shown in Fig. \ref{fig4}. The peculiarity of this case is the
existence of density spikes near the trap edges~\cite{roccuzzo2021}.
The ground state with a density peak (drop) at $x=0$ emerges if the
trap can accommodate an odd (even) number $m$ of supersolid periods
$2 h = m \lambda$.
Sudden reduction of the repulsive dipolar interactions gives rise to
flow of superfluid matter into the central region of the trap, which
shows up as raising (and subsequent oscillations) of the central
density peaks (see Fig.~\ref{fig4}b). This is a clear manifestation
of the key property of supersolids, namely free flow of the
superfluid component through its crystalline phase.

To identify the frequencies of particular modes we perform Fourier
analysis of the time-dependent width $W(t)$ calculated according to
Eq.(\ref{W}), with $\psi(x,t)$  being the numerical solution of the
eGPE (\ref{gpe}). The Fourier transform reveals three frequencies
(see Fig. \ref{fig4}d). The lower frequency is associated with the
superfluid component, while other two higher frequencies are linked
to the crystalline component of the SS. The additional frequency
originates from two pronounced density peaks (edge states) of
dipolar BEC confined to a box-potential (see Fig. \ref{fig4}a).
These density peaks can be viewed as a separate crystal component,
oscillating at a different frequency than the main crystalline
phase, occupying the central part of the box-trap.

\subsection{Supersolid in a parabolic potential}

\begin{figure}[htb]
\centerline{{\large a)} \hspace{6cm} {\large b)}}
\centerline{\includegraphics[width=6cm,height=4cm,clip]{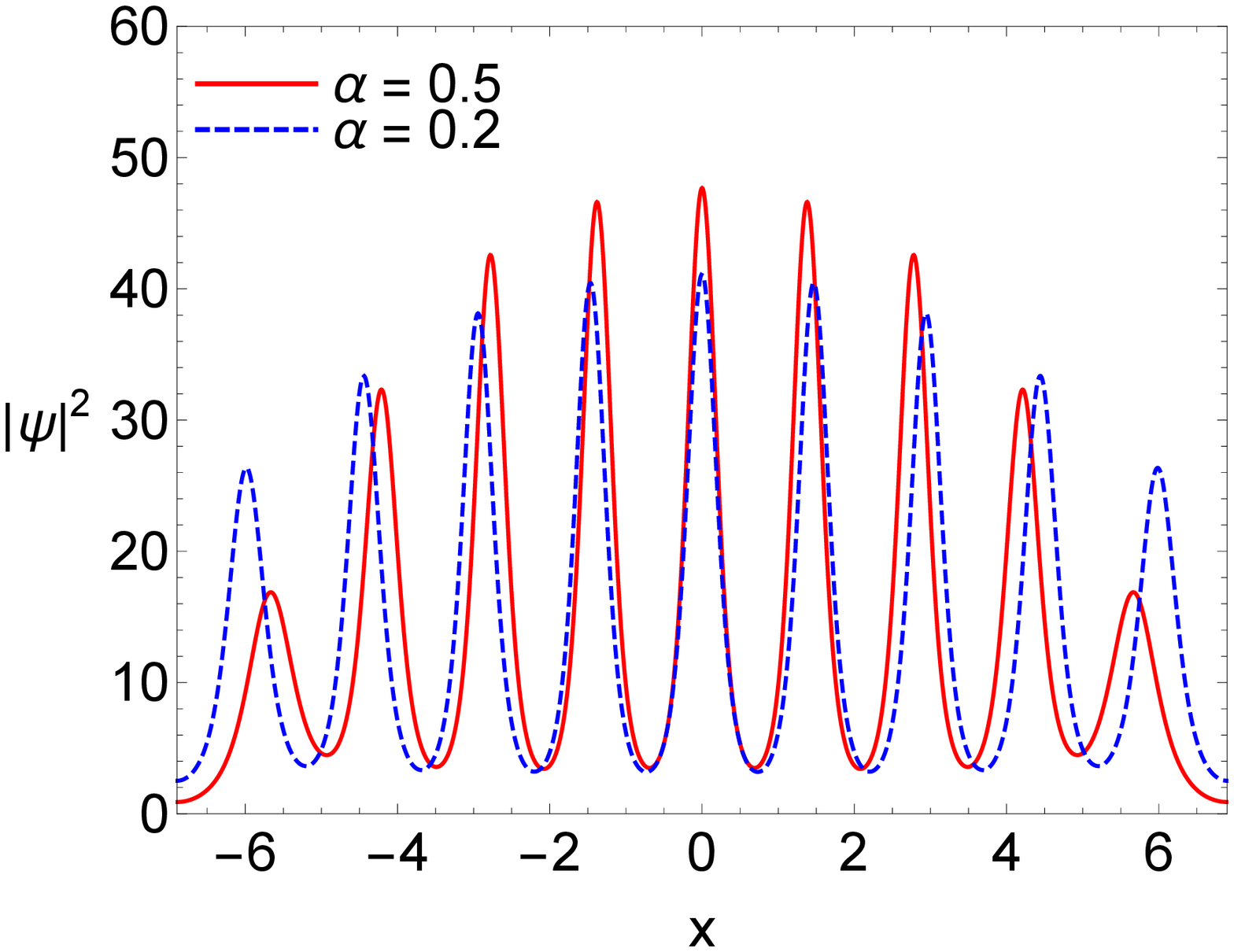}\qquad
            \includegraphics[width=6cm,height=4cm,clip]{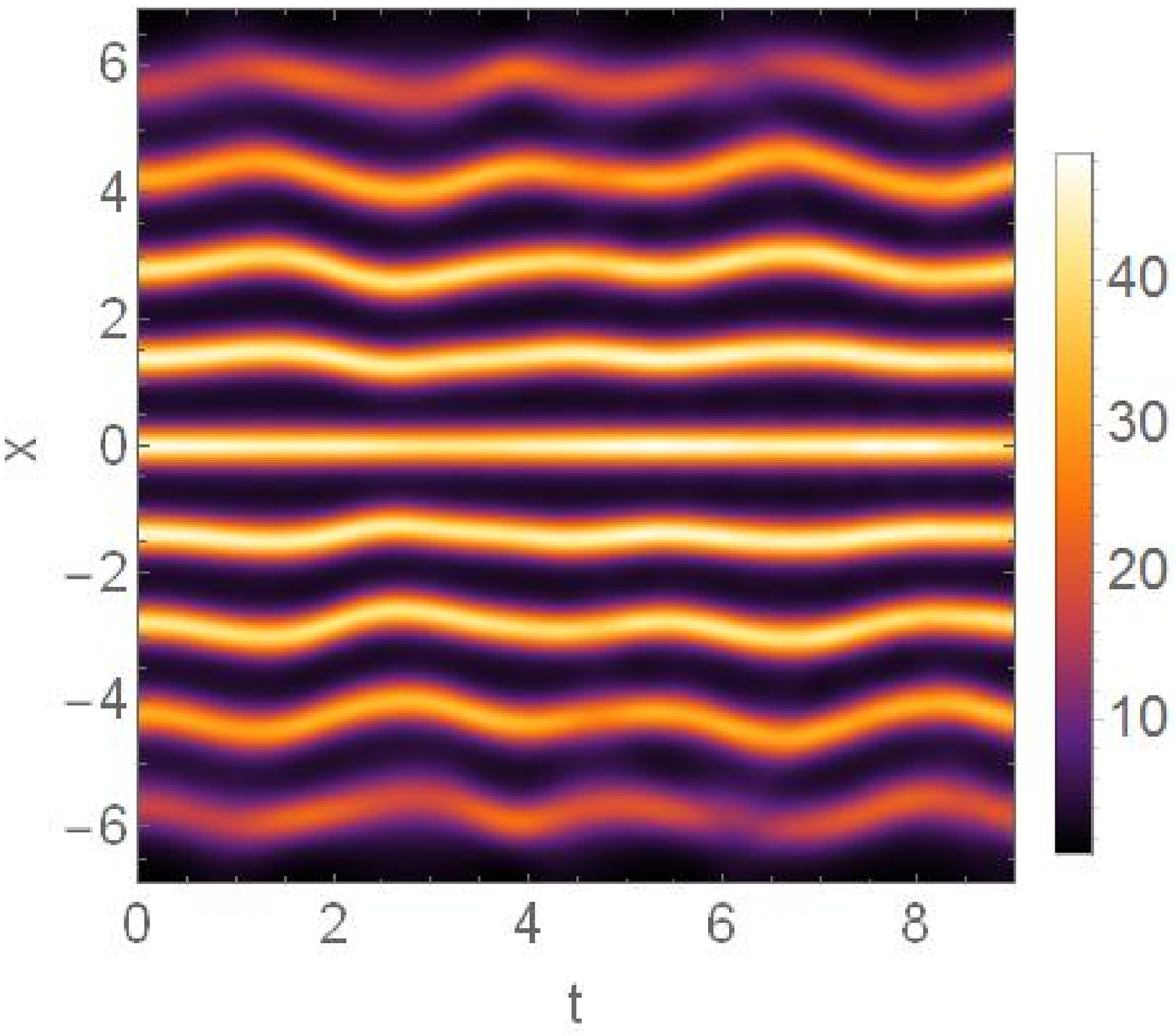}}
\centerline{{\large c)} \hspace{6cm} {\large d)}}
\centerline{\includegraphics[width=6cm,height=4cm,clip]{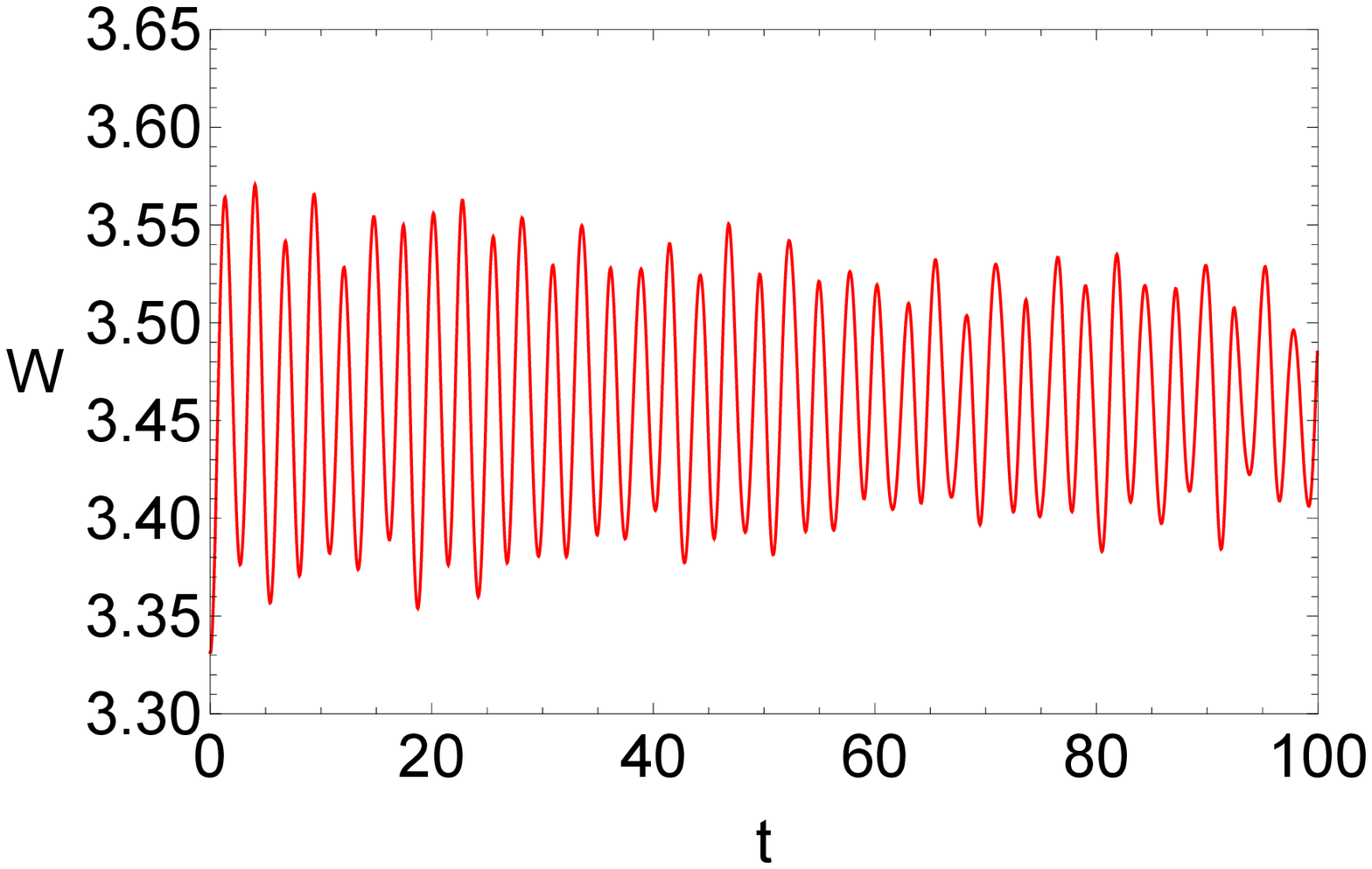}\quad
            \includegraphics[width=6cm,height=4cm,clip]{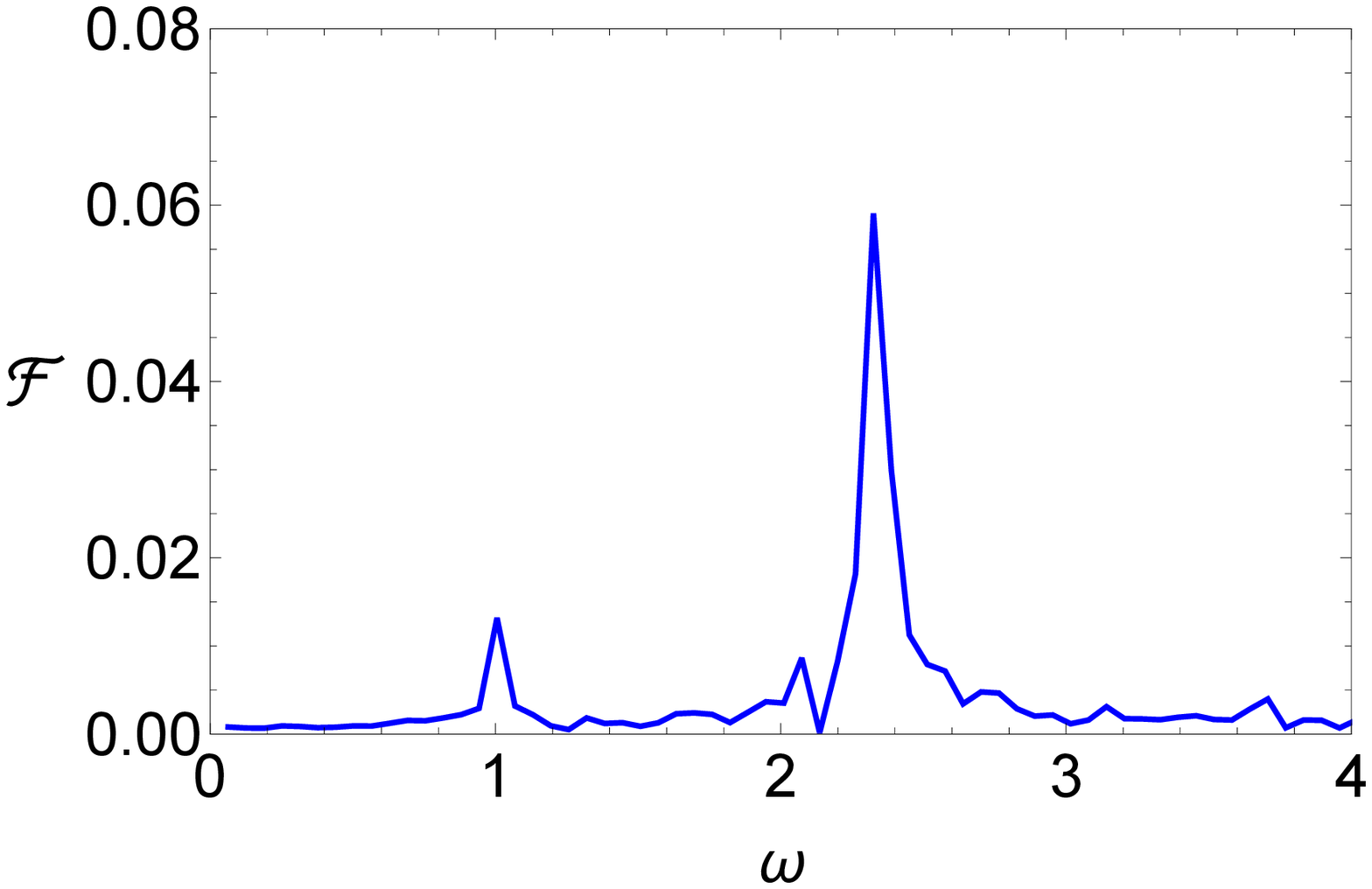}}
\caption{a) The ground state of the supersolid in the parabolic trap
$V(x)=\alpha \, x^2$ for $\alpha = 0.5$ (red solid line) and $\alpha
= 0.2$ (blue dashed line). b) The density plot, showing the
collective oscillations of the supersolid induced by reducing the
strength from $\alpha = 0.5$ to $\alpha = 0.4$. c) The average width
of the supersolid $W(t)$ shows complex dynamics, giving evidence of
the superposition of few oscillations. d) The Fourier transform of
the effective width ${\cal F}({\rm W})$ reveals two main
frequencies, corresponding to oscillations of the superfluid
component ($\omega_s \sim 1$) and crystalline phase ($\omega_c \sim
2.3$), respectively. Parameter values: $g=1$, $p=-0.006$,
$\gamma=-0.001$. The length of integration domain is $L=9\lambda$.}
\label{fig5}
\end{figure}

The parabolic potential is a frequently used setting in experiments
with ultra-cold quantum gases. The ground state of the supersolid
confined to a parabolic trap, numerically produced by the Pitaevskii
phenomenological damping procedure, is depicted in Fig. \ref{fig5}.
The figure gives evidence that the stronger is the parabolic trap,
the more amount of superfluid matter is expelled to the trap
periphery. A similar effect was noticed for the box trap where the
edge states with increased matter density emerge even for parameter
regimes, where the SS phase is absent. As a consequence, the lateral
density peaks in the parabolic trap become less pronounced.

To study the collective dynamics of the SS in a parabolic trap we
introduce its ground state (red solid line in Fig. \ref{fig5}a) into
the eGPE as initial condition and reduce the trap strength from
$\alpha=0.5$ to $\alpha=0.4$. The resulting collective dynamics of
the supersolid is shown through the density plot in Fig.
\ref{fig5}b. Notable periodic variation of the inter-peak separation
can be seen. The average width of the supersolid $W(t)$ also shows
complex dynamics, giving evidence of the superposition of few
oscillations (Fig. \ref{fig5}c). The Fourier transform of the
effective width ${\cal F}(W)$ reveals two main frequencies
(Fig.~\ref{fig5}d), corresponding to oscillations of the superfluid
component ($\omega_s \sim 1$) and crystalline phase ($\omega_c \sim
2.3$), respectively.

\begin{figure}[htb]
\centerline{{\large a)} \hspace{6cm} {\large b)}}
\centerline{\includegraphics[width=6cm,height=4cm,clip]{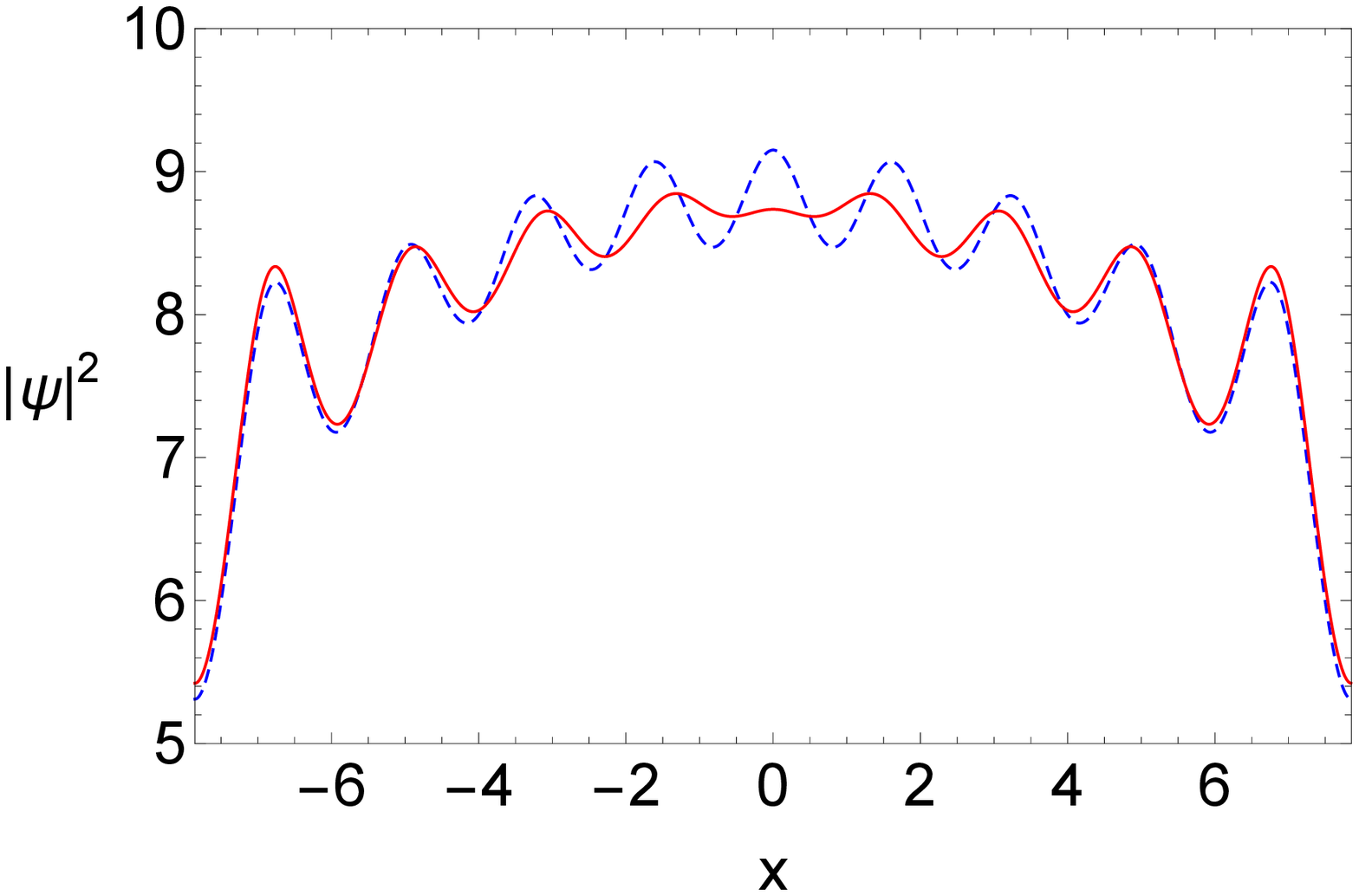}\qquad
            \includegraphics[width=6cm,height=4cm,clip]{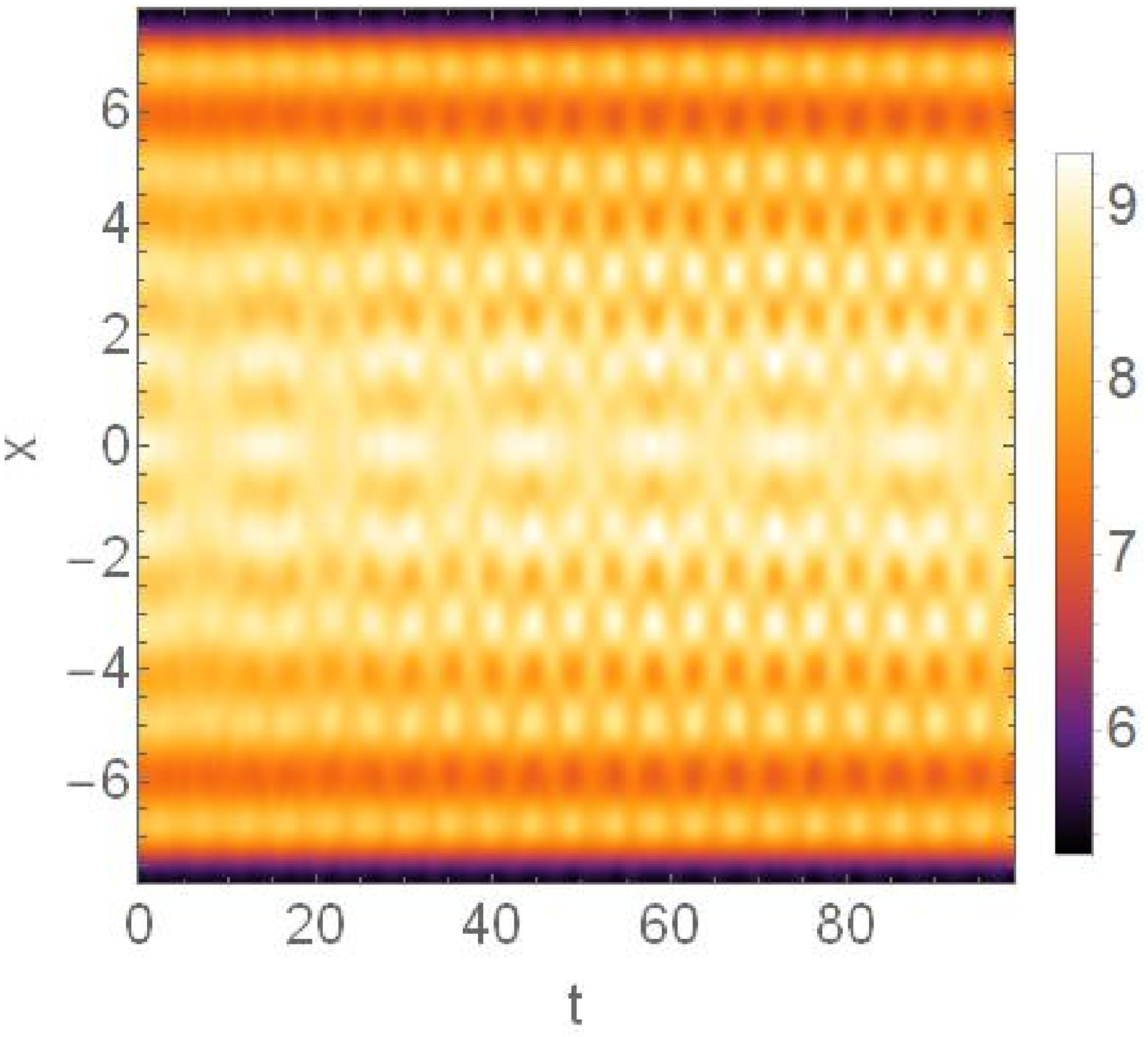}}
\centerline{{\large c)} \hspace{6cm} {\large d)}}
\centerline{\includegraphics[width=6cm,height=4cm,clip]{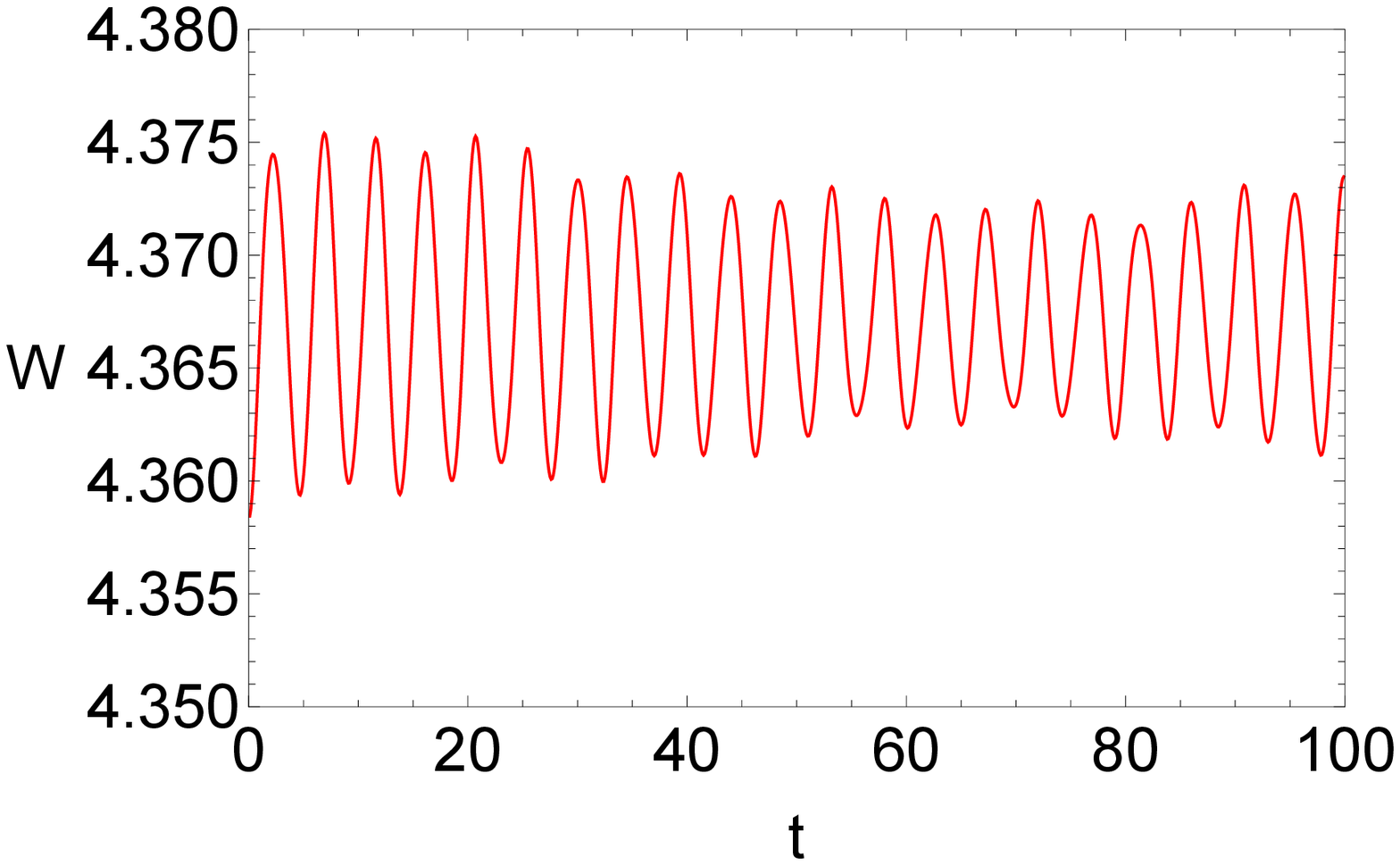}\quad
            \includegraphics[width=6cm,height=4cm,clip]{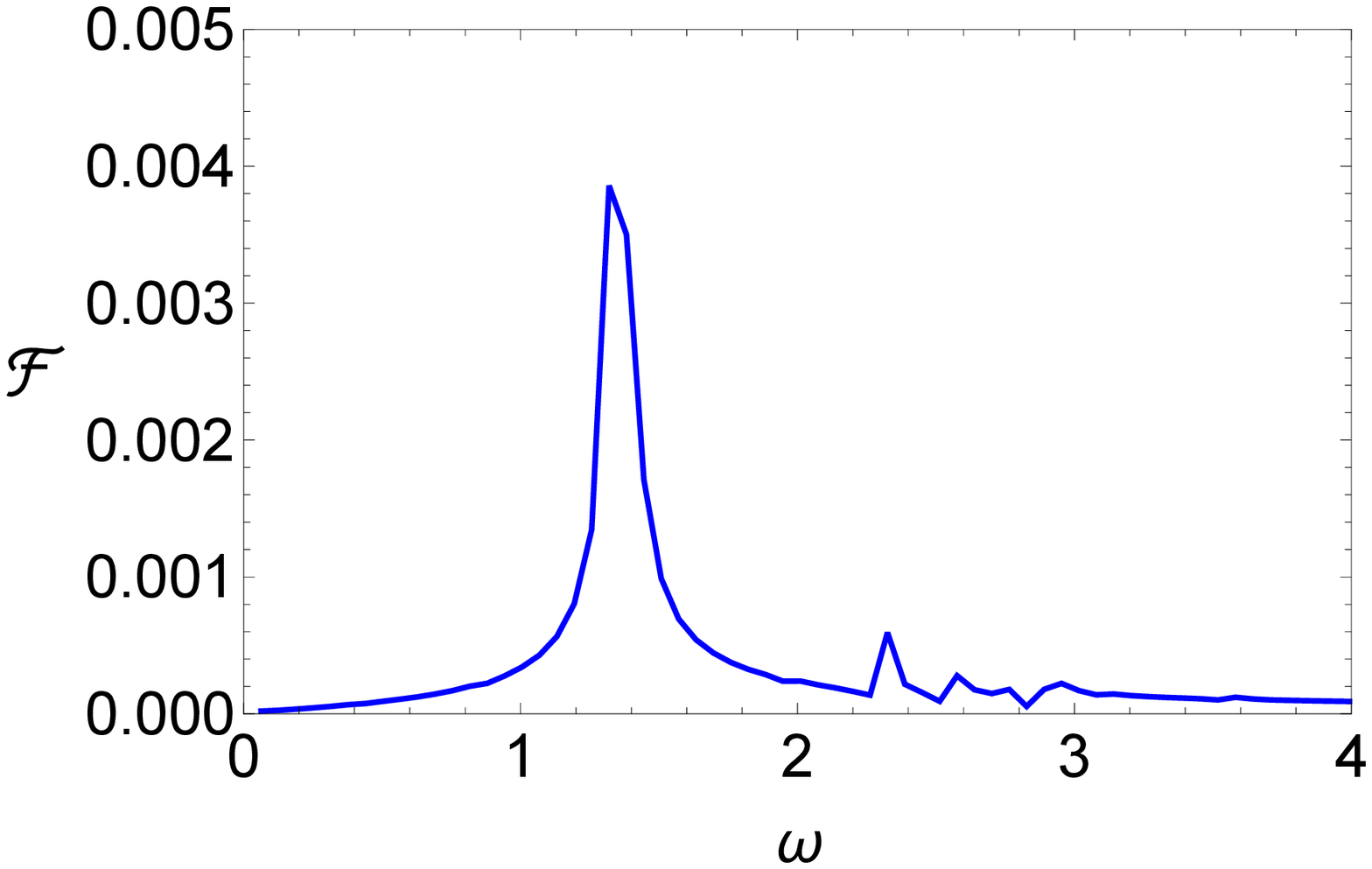}}
\caption{a) The ground state density profile of the dipolar BEC
confined to a weak parabolic potential $V(x)=\alpha \, x^2$ for
$\alpha = 0.05$ (blue dashed line). When the parameter is reduced to
$\alpha = 0.048$ the condensate starts to oscillate, whose density
profile is shown for $t=7$ (red solid line). b) The density of the
condensate as a function of time and space followed by reduction of
the parameter $\alpha$. c) The average width of the condensate
$W(t)$ performs oscillations with slowly varying amplitude. d) The
Fourier transform ${\cal F}({\rm W})$ reveals two main frequencies,
corresponding to oscillations of the superfluid ($\omega_s \simeq
1.4$) and crystalline ($\omega_c \simeq 2.4$) components,
respectively. Parameter values: $g=1$, $g_{dd}=-1.3$, $p=-0.005$,
$\gamma=-0.002$, $k_r=3.6$, $\lambda=1.75$, $L=9\lambda$. }
\label{fig6}
\end{figure}

Figure \ref{fig5} represents the example of SS with dominating
crystalline component, which shows a stronger peak in the Fourier
spectrum than that of the superfluid component. Now we consider the
opposite situation with dominating superfluid phase, which is shown
in Fig. \ref{fig6}. The ground state is numerically obtained by the
Pitaevskii phenomenological damping procedure starting from the
uniform condensate with linear density $n_0=A^2=8.128$. For selected
parameter values the density modulations are less pronounced in the
central region of the trap, while notable density peaks show up near
the borders, similar to edge states in the box-trap (see Fig.
\ref{fig4}a).
To induce oscillations of the condensate initially residing in the
ground state of the parabolic trap, we slightly change its strength
$\alpha$. Figure \ref{fig6}\,c illustrates the variation of the
effective width of the condensate caused by a 5 \% reduction of the
parameter $\alpha$. Here a superposition of two frequencies can be
noticed. The Fourier transform shown in Fig. \ref{fig6}\,d reveals
these two frequencies, associated with the superfluid and
crystalline components. The crystalline phase, although quite
weakly, shows up through the density peaks near trap edges. By
comparing the results shown in figures \ref{fig5} and \ref{fig6} we
conclude that when the SS is perturbed, its crystalline (superfluid)
component oscillates with higher (lower) frequency, in accordance
with experimental findings \cite{tanzi2019b}.

\subsection{Estimation of parameter values}

Now we estimate the parameters of our model using the quantities
relevant to experiments \cite{tanzi2019b,guo2019}. Consider
$^{162}$Dy atoms, whose magnetic dipole moment, background $s$-wave
scattering length and atomic mass are $\mu_{Dy}= 10 \,\mu_{B}$,
$a_{bg}=-130 \, a_0$, $m_{Dy}=2.7 \times 10^{-25}$ kg, respectively,
with $\mu_{B}$, $a_0$ being the Bohr magneton and Bohr radius. The
roton instability sets in, thus supersolid phase emerges, when the
$s$-wave scattering length is tuned to $a_s = -98 a_0 \simeq 0.005
\mu$m. The condensate of $N=4 \times 10^4$ atoms is held in a tight
quasi-1D trap with radial confinement frequency $\omega_{\bot}=2 \pi
\times 110$~Hz. The corresponding radial harmonic oscillator length,
which is the adopted length scale in this work, is $a_{\bot}=0.75 \,
\mu$m. The unit of time is $\omega_{\bot}^{-1}=1.4 \times 10^{-3}$
s. The ratio between the strengths of dipolar and contact
interactions, computed with above defined parameters, is found to be
$\epsilon_d=\mu_0 \mu_{Dy}^2 m_{Dy}/12\pi \hbar^2 a_s \simeq 1.34$,
therefore we have the dipolar interaction dominated regime. The
coefficients of two-body contact interactions and long-range dipolar
interactions in physical units are $g_0=4\pi \hbar^2
a_s/m_{Dy}=-2.68 \times 10^{-51} \, {\rm kg} \cdot {\rm m}^5/{\rm
s}^2$ and $C_d = \mu_0 \mu_{Dy}^2 = 1.08 \times 10^{-50} \, {\rm kg}
\cdot {\rm m}^5/{\rm s}^2$, yielding the dimensionless parameter
$g_{dd}=C_d/g_0 \simeq - 4$. The strength of three-body atomic
interactions $K_3=K_r+i\, K_{i}$ in dysprosium condensate was
reported to be in the range $K_r=(3.91 - 19.6) \times 10^{-39} \,
\hbar \cdot {\rm m}^6/{\rm s}$ for real conservative part and
$K_{i}=7.8 \times 10^{-42} \, \hbar \cdot {\rm m}^6/{\rm s}$ for
imaginary part, characterizing the three-body recombination rate
\cite{bisset2015}. Since the imaginary part of this parameter is
smaller than its real part by three orders of magnitude, we have
omitted it in the eGPE (\ref{gpe}), leaving only the conservative
part $g_2=K_r$. The coefficient of quintic nonlinearity, computed
using this value and above defined parameters is found to be in the
range $p = - (0.007 - 0.036)$. The coefficient of quantum
fluctuations corresponding to above mentioned parameter values is
estimated as $\gamma \sim  - 0.05$.

Numerical simulations are performed using the integration domain of
length ${\cal L}=(9 - 12)\lambda \, a_{\bot} \simeq (10 - 14) \,
\mu$m. The amplitude of the background wave in dimensionless units
is defined as $A=\sqrt{2\,|a_s|\,N/{\cal L}} \simeq 6$, where $N$
stands for the number of atoms in the condensate. The linear density
of the condensate is $n_l=N/{\cal L} \sim 4 \times 10^{3} \mu
m^{-1}$.

The above estimates for dimensionless parameters obtained from
experimentally possible quantities are in the range of values used
in our numerical simulations. Since the radial confinement length is
much greater than the atomic scattering length ($a_{\bot} \gg a_s$),
usage of parameter values relevant to 3D geometry for numerical
simulations in quasi-1D setting is justified. Some deviations can be
adjusted by changing the tunable parameters of the system, such as
the $s$-wave scattering length, frequency of radial confinement, and
the strength of dipolar interactions. As for the roles of the
repulsive three-body atomic collisions and quantum fluctuations,
these are two possible factors responsible for the existence of
stable supersolids \cite{lu2015,tanzi2019b}, and for a complete
description of the problem, both terms should be taken into account.

\section{Conclusions}

We have explored the static and dynamic properties of a supersolid
state in quasi-one-dimensional dipolar quantum gases in three
different settings, including the periodic boundary conditions,
box-like confinement, and parabolic potential. The model is based on
the extended Gross-Pitaevskii equation involving the effect of
quantum fluctuations and three-body atomic interactions, which
correspond to quartic and quintic nonlinearities, respectively. The
dipolar interatomic repulsion gives rise to density peaks near the
edges of the confining potential, which is shown for the box-like
and parabolic traps. The analytic approach, based on minimization of
the Gross-Pitaevskii energy functional is developed to estimate the
amplitude of the supersolid, which is linked to the experimentally
relevant quantity called the density contrast. The supersolid states
with weak and strong modulation of the condensate density are
produced by numerical methods. The amplitude of the weakly modulated
supersolid well agrees with the theoretical prediction. The
oscillations of the supersolid residing in the ground state are
induced by changing the strength of atomic interactions, and
variation of the confining potential. By recording the
time-dependent effective width of the perturbed condensate and
subsequent Fourier analysis of the obtained data we identify two
distinct oscillation frequencies, associated with the superfluid and
crystalline components. These frequencies depend on the type and
strength of the external potential. The edge states of a dipolar
condensate in the confining potential can produce additional
frequency in the Fourier spectrum. In numerical simulations we
observed the key property of the supersolid - free flow of the
superfluid fraction through the crystalline component of the
condensate.

\section*{Acknowledgments}

We thank Dr. E. N. Tsoy for valuable discussions. The work of BBB is
partially supported by the Ministry of Innovative Development of
Uzbekistan through International Uzbekistan-Turkey Project
UT-$\Phi$A-2020-3.

\end{document}